\documentclass[aip, jcp 
amsmath,amssymb,
reprint,%
]{revtex4-2}

\usepackage{graphicx}
\usepackage{dcolumn}
\usepackage{bm}

\usepackage[utf8]{inputenc}
\usepackage[T1]{fontenc}
\usepackage{mathptmx}
\usepackage{etoolbox}
\usepackage{color}

\usepackage{amsmath,amsthm,amssymb, graphicx, array}

\usepackage[percent]{overpic}

\usepackage[caption=false]{subfig} 

\newcommand{\Lfn}[1]{{\cal L}^{(#1)}}
\newcommand{\myT}{{\cal T}}
\graphicspath{{./Figures/}}
		\newcommand{\Ltilde}[1]{{\cal \widetilde{L}}^{(#1)}}
		
\makeatletter
\def\@email#1#2{%
 \endgroup
 \patchcmd{\titleblock@produce}
 {\frontmatter@RRAPformat}
 {\frontmatter@RRAPformat{\produce@RRAP{*#1\href{mailto:#2}{#2}}}\frontmatter@RRAPformat}
 {}{}
}%
\makeatother

\begin{document}

\preprint{AIP/123-QED}


\title{When Dephasing Fails: Thermodynamic Consequences of Decoherence Models in Quantum Transport}

\author{E. Erdogan}%
\affiliation{Department of Physics, Illinois State University, Normal, Illinois 61790, United States; Department
	of Chemistry, Illinois State University, Normal, Illinois 61790}
	
\author{J. P. Bergfield* }
\email{jpbergf@ilstu.edu}
\affiliation{Department of Physics, Illinois State University, Normal, Illinois 61790, United States; Department
of Chemistry, Illinois State University, Normal, Illinois 61790}


\date{\today}

\begin{abstract}
Understanding how decoherence influences heat and information flow is essential for realizing the promise of quantum technologies. Two widely used models for incorporating decoherence in quantum transport are the voltage probe (VP), which imposes local charge current conservation, and the voltage–temperature probe (VTP), which also conserves heat current. 
Although these models are often treated as functionally equivalent, we demonstrate that this equivalence actually exists only 
under highly symmetric conditions, which may be challenging to achieve experimentally.
Under asymmetric coupling or thermal bias, the VTP respects thermodynamic constraints and enforces decoherence in both charge and heat channels, while the VP instead acts as a source or sink of heat. Strikingly, the VP can fail to model decoherence in the heat transport entirely, even with large probe coupling strengths. 
Using a benzene-based molecular junction as a realistic example, we show that these effects significantly impact the predicted heat transport.
These results establish that the VP and VTP models are not interchangeable: only the VTP provides a thermodynamically consistent framework for modeling decoherence in quantum transport.
\end{abstract}

\maketitle

\section{Introduction}

As Schr\"odinger famously remarked, entanglement is ``not {\em one}, but rather \emph{the} characteristic trait of quantum mechanics, the feature that enforces its complete departure from classical lines of thought''.\cite{Schrödinger_1935}
While entanglement refers to certain nonclassical correlations enabled by quantum coherence, the underlying phase coherence itself plays a broader and more fundamental role across quantum systems.
For example, in coherent conductors, transport depends not only on available energy levels but also on the phase relationships among distinct transport paths.\cite{evers2020advances} One of the most striking consequences of quantum interference (QI) is the formation of transmission nodes, features in the transport spectrum stemming from the complete destructive interference between accessible transport amplitudes. 
In the vicinity of a node, heat and charge are affected differently, enabling novel energy conversion and control mechanisms that are inaccessible to classical systems.
These interference-enhanced effects have been exploited across a range of applications, including scalable thermoelectric materials,\cite{bergfield2009thermoelectric,bergfield2010giant,bennett2024quantum,Finch09} tunable entropy and heat flow,\cite{shastry2020scanning,bergfield2014thermoelectric,bergfield2015tunable,bergfield2013probing,majidi2024heat}, molecular switches,\cite{cardamone2006controlling,stafford2007quantum,ke2008quantum,chen2024quantum} and low-loss dielectric materials.\cite{bergfield2015harnessing}

QI effects on the electronic transport can be remarkably robust, persisting even at room temperature and in noisy environments for certain molecular and mesoscopic systems.\cite{aradhya2012dissecting,arroyo2013signatures,guedon2012observation,markussen2014temperature,liu2018quantum,yang2018quantum,bergfield2024identifying} However, QI remains fundamentally susceptible to dephasing and decoherence induced by interactions with additional degrees of freedom, such as vibrational modes, electromagnetic fluctuations, etc. A powerful and widely adopted theoretical framework for incorporating decoherence into quantum transport was introduced by B\"uttiker, who showed that fictitious terminals (probes) can mimic phase-randomizing processes while preserving global conservation laws.\cite{buttiker1986four,buttiker1988coherent} Building on earlier work by Engquist and Anderson,\cite{engquist1981definition} B\"uttiker formalized the concept of the \emph{voltage probe} (VP), defined as a terminal with vanishing net charge current. Such probes absorb and re-emit electrons stochastically, effectively randomizing their phase and simulating decoherence without the need to specify the microscopic details. 

While this formalism is both elegant and broadly applicable,\cite{Guo2012,Chen2014,Nozaki2008,Nozaki2012,kilgour2015charge,kilgour2018coherence} it was originally formulated in the zero-temperature limit, where charge conservation alone suffices to define the probe potential unambiguously.
At finite temperatures, however, thermoelectric effects introduce additional terms which render charge conservation alone insufficient as the only condition needed to ensure local thermodynamic equilibrium.\cite{buttiker1989chemical,bergfield2014thermoelectric}
Although B\"uttiker anticipated these issues,\cite{buttiker1989chemical} a fully consistent treatment was only later developed with the {\em voltage–temperature probe} (VTP) formalism,\cite{bergfield2013probing,meair2014local} in which both charge and heat currents into the probe vanish, enabling general operational definitions of the local temperature and chemical potential.\cite{Stafford2016local,Shastry2015local}
%

VPs and VTPs both model decoherence in quantum transport; however, they impose fundamentally different thermodynamic constraints.
A VP dynamically adjusts its chemical potential to eliminate net charge flow, thereby defining a local electrochemical potential.
By contrast, a VTP additionally adjusts its temperature to eliminate net heat flow, enforcing local thermodynamic equilibrium.
Although it is well established that the chemical potential and temperature of a probe can differ significantly depending on the choice of probe model, particularly in systems with substantial thermoelectric effects,\cite{bergfield2013probing,bergfield2014thermoelectric, bergfield2015tunable} the degree to which these differences affect the transport through the other physical leads remains unclear.

Here, we investigate how the choice of probe affects the heat and charge transport in quantum-coherent circuits. We derive analytic expressions for linear-response currents in both electronic and purely thermal configurations, allowing a direct comparison of the physical mechanisms underlying each model. 
While VP and VTP probes may appear similar under symmetric conditions, we find they are generally \emph{thermodynamically inequivalent}, with differences that become especially pronounced under asymmetries in tunnel coupling or thermal bias.
Finally, we calculate the transport through a representative molecular conductor, the meta-connected Au-1,3-benzenedithiol-Au (1,3-BDT) junction, to demonstrate the magnitude of these effects in a realistic system.

\section{Quantum Transport Theory}

We investigate charge and heat (entropy) transport in interacting open quantum systems composed of a nano-system, such as a molecule, coupled to \( M \) macroscopic electrodes, modeled as ideal Fermi gases. To describe these systems, we employ the non-equilibrium Green’s function (NEGF) formalism, which offers a rigorous and general framework for computing transport in open quantum systems.\cite{HaugAndJauhoBook,stefanucci2013nonequilibrium}.

We focus on systems in which transport is predominantly elastic and phase coherent. In linear-response, the electrical current $-eI^{(0)}_\alpha$ and heat current $I^{(1)}_\alpha$ flowing into reservoir $\alpha$ of such a system may be expressed as\cite{meir1992landauer,buttiker1985generalized,sivan1986multichannel,bergfield2009thermoelectric}
\begin{equation}
	\label{eq:linear_response}
	I_\alpha^{(\nu)} = \sum_{\beta=1}^{M} \left[ {\cal L}^{(\nu)}_{\alpha\beta} (\mu_\beta - \mu_\alpha) + \Lfn{\nu+1}_{\alpha\beta} \frac{T_\beta - T_\alpha}{T_0} \right],
\end{equation}
where the Onsager functions are given by
\begin{equation}
	\Lfn{\nu}_{\alpha\beta} = \frac{1}{h} \int dE\, (E - \mu_0)^\nu\, \myT_{\alpha\beta}(E) \left(-\frac{\partial f_0}{\partial E} \right),
\end{equation}
and $f_0(E) = \left[ \exp\left( (E - \mu_0)/k_B T_0 \right) + 1 \right]^{-1}$ is the equilibrium Fermi-Dirac distribution of the electrodes at chemical potential $\mu_0$ and temperature $T_0$.
The transmission function may be expressed as 
\begin{equation}
{\myT}_{\alpha\beta}(E)={\rm Tr}\left\{ {\bm \Gamma}^\alpha(E) {\cal G}(E) {\bm \Gamma}^\beta(E) {\cal G}^\dagger(E)\right\},
\label{eq:transmission_prob}
\end{equation} 
where ${\bm \Gamma}^\alpha(E)$ is the tunneling-width matrix for lead $\alpha$
and ${\cal G}(E)$ is the retarded Green's function of the junction. 

\subsection{Electronic Structure Theory}


As an example, we simulate the transport through a Au–1,3-benzenedithiol–Au (1,3-BDT) molecular junction,\cite{bergfield2009many,cardamone2006controlling,solomon2008understanding,markussen2010relation,markussen2011graphical,bennett2024quantum} in which a benzene ring is covalently anchored to gold electrodes via thiol linkers. The meta connectivity of 1,3-BDT gives rise to destructive QI, resulting in a mid-gap transmission node.\cite{pedersen2014quantum,bergfield2010coherent,solomon2008understanding,li2015mechanical,bennett2024quantum} Although in general interference features originate from the structure of the many-body Hilbert space,\cite{bergfield2011novel,barr2013transmission} they can be captured accurately in this class of systems by effective single-particle models. Accordingly, we employ a H\"uckel description, which reproduces the essential interference signatures characteristic of benzene-based molecular junctions.\cite{cardamone2006controlling,solomon2008quantum,markussen2010relation,liu2018quantum}

We define the molecular Hamiltonian as
\begin{equation}
	H_{\rm mol} = \sum_n \varepsilon_n d^\dagger_n d_n + \sum_{\langle ij \rangle} t_{ij} d^\dagger_i d_j + \text{H.c.},
\end{equation}
where \( d^\dagger_n \) and \( d_n \) respectively create and annihilate an electron on site \( n \), and \( t_{ij} = 2.7 \, \mathrm{eV} \) denotes the nearest-neighbor hopping integral. These parameters are extracted from gas-phase spectroscopy using an effective field theory for the conjugated $\pi$-system.\cite{barr2012effective} 
The influence of thiol anchor groups, which formally break particle-hole symmetry and perturb level alignment, can be incorporated via first-principles corrections to the on-site energies.\cite{barr2012effective,bergfield2024identifying,bennett2024quantum} However, since their effect is typically small near a mid-gap node\cite{bennett2024quantum} we have omitted them here for clarity.

%
%
In this theory, the junction Green’s function is given by
\begin{equation}
	\mathcal{G}(E) = \left( E \mathbf{1} - H_{\rm mol} - \Sigma_{\rm T} \right)^{-1},
\end{equation}
where \( \Sigma_{\rm T} = -\tfrac{i}{2} \sum_{\alpha=1}^{M} {\bm \Gamma}_\alpha \) is the total tunneling self-energy due to all electrodes. The tunneling-width matrix associated with electrode \( \alpha \) is defined as
\begin{equation}
	\left[{\bm \Gamma}_\alpha\right]_{nm} = 2\pi \sum_{k \in \alpha} V_{nk} V_{mk}^* \delta(E - \epsilon_k),
\end{equation}
where \( V_{nk} \) denotes the coupling amplitude between molecular orbital \( n \) and electrode state \( k \), and \( \epsilon_k \) is the energy of state \( k \). Throughout this work, we adopt the wide-band limit, in which each coupling matrix \( {\bm \Gamma}_\alpha \) is taken to be energy-independent.

\subsection{Entropy Production}

In addition to the currents, thermodynamic consistency requires that any steady-state transport process produces non-negative entropy, in accordance with the second law. In the linear-response regime, and assuming local equilibrium in each reservoir, the total rate of entropy production can be expressed as the sum of entropy currents entering the terminals:
\begin{equation}
	\dot{S}_{\rm tot} = \sum_\alpha \frac{I_\alpha^{(1)}}{T_\alpha},
	\label{eq:entropy_production}
\end{equation}
where $I_\alpha^{(1)} / T_\alpha $ is the entropy current flowing into terminal \( \alpha \), and \( T_\alpha \) is its temperature. This expression derives from the first and second laws applied to open quantum systems coupled to equilibrated reservoirs and provides a rigorous criterion for thermodynamic admissibility in steady state.

A strictly positive \( \dot{S}_{\rm tot} \) indicates irreversible processes such as scattering, thermal gradients, or coupling to an incoherent environment. In contrast, \( \dot{S}_{\rm tot} = 0 \) implies a reversible process, e.g., fully coherent ballistic transport with no net entropy generation. Crucially, in quantum-coherent systems, entropy production does not originate from the unitary evolution of the isolated system, but from its coupling to external environments. These environments, modeled here via fictitious terminals (probes), serve to simulate decoherence and inelastic effects under controlled constraints.

The thermodynamic behavior of the system depends sensitively on the type of probe employed. A VTP, which adjusts both its chemical potential and temperature to ensure zero net particle and heat flow, represents an idealized thermalizing environment. By construction, it neither injects nor extracts entropy, and so contributes nothing directly to \( \dot{S}_{\rm tot} \); any entropy production reflects intrinsic dissipation among the physical terminals. A VP, by contrast, enforces only charge conservation while maintaining an externally imposed temperature. As such, it can become a non-equilibrated reservoir, i.e. an external entropy source or sink.

\section{Three-Terminal Circuits}

To investigate the influence of probe model on the coherent transport, we derive analytic expressions for the electrical and thermal response of three-terminal quantum junctions composed of two macroscopic electrodes (left, $\mathrm{L}$ and right, $\mathrm{R}$) and a third probe terminal ($\mathrm{P}$) which introduces dephasing and is modeled by either a VP or VTP. 
In both models, the probe satisfies the charge conservation condition \( I_{\rm P}^{(0)} = 0 \), dynamically adjusting its chemical potential \( \mu_{\rm P} \) to eliminate net charge flow. The VTP model imposes an additional constraint of local thermal equilibrium by also requiring the heat current to vanish, \( I_{\rm P}^{(1)} = 0 \), which necessitates a self-consistent determination of the probe temperature \( T_{\rm P} \). Although each probe is characterized by a single electrochemical potential and temperature, it may couple locally to a single site or nonlocally to multiple orbitals.\cite{bergfield2010coherent} 

In the sections that follow, we consider two representative configurations: (i) an \textit{electronic circuit}, in which charge and heat transport are driven by voltage and temperature differences between the $\mathrm{L}$ and $\mathrm{R}$ terminals; and (ii) a \textit{pure thermal circuit}, in which a temperature gradient is applied while the chemical potentials of the leads are adjusted to enforce open-circuit conditions, \( I^{(0)}_\alpha = 0 \forall \alpha \). By comparing the VP and VTP models under each biasing scheme, we establish a foundation for exploring how probe constraints shape measurable transport observables under various symmetry conditions.


\subsection{Electronic Circuits}

In linear response, the condition \( I_{\rm P}^{(0)} = 0 \) leads to a general expression for the probe chemical potential\cite{buttiker1986four,buttiker1986role}
\begin{equation}
	\mu_{\rm P} = \frac{1}{ -{\cal L}^{(0)}_{\rm PP} } \sum_{\alpha \neq P} \left[ {\cal L}^{(0)}_{{\rm P} \alpha} \mu_\alpha + {\cal L}^{(1)}_{{\rm P}\alpha} \frac{T_\alpha - T_{\rm P}}{T_0} \right],
	\label{eq:muP_vp1}
\end{equation}
where \( T_0 \) is the reference temperature and \( T_{\rm P} \) depends on the probe type: fixed in the VP model, but determined self-consistently for the VTP case. Gauge invariance and current conservation impose the sum rules
\begin{equation}
	\sum_\alpha {\cal L}^{(\nu)}_{\alpha\beta} = \sum_\beta {\cal L}^{(\nu)}_{\alpha\beta} = 0,
	\label{eq:gauge_invariance}
\end{equation}
from which it follows that \( \mathcal{L}^{(\nu)}_{\rm PP} = -\sum_{\alpha \neq P} \mathcal{L}^{(\nu)}_{{\rm P}\alpha} \). This identity simplifies Eq.~\eqref{eq:muP_vp1} and underscores the internal consistency of the effective transport description.

When the third terminal acts as a VP, we insert the value of \( \mu_{\rm P} \) into the terminal current expressions, yielding
\begin{align}
	I_{\rm R, VP}^{(\nu)} &= \Ltilde{\nu,\nu}_{\rm RL} (\mu_{\rm L} - \mu_{\rm R}) \nonumber \\
 & +\left[ \Ltilde{\nu+1,\nu}_{\rm RL} \frac{T_{\rm L} - T_{\rm R}}{T_0} + \Ltilde{\nu+1,\nu}_{\rm RP} \frac{T_{\rm P}^{\rm VP} - T_{\rm R}}{T_0} \right],
	\label{eq:IRVP}
\end{align}
where we have introduced the effective Onsager coefficients using the Schur complement
\begin{equation}
	\Ltilde{\nu,\xi}_{\alpha\beta} = \Lfn{\nu}_{\alpha\beta} + \frac{\Lfn{\xi}_{\alpha P} \Lfn{\nu-\xi}_{P \beta}}{- \Lfn{0}_{\rm PP}}.
	\label{eq:effective_onsager}
\end{equation}
These coefficients preserve the symmetry and conservation laws of the original Onsager matrix,
\begin{equation}
	\sum_\alpha \Ltilde{\nu,\xi}_{\alpha\beta} = \sum_\beta \Ltilde{\nu,\xi}_{\alpha\beta} = 0.
  \label{eq:effOnsager_gauge_invariance}
\end{equation}

For the VTP model, both charge and heat conservation are enforced. The chemical potential \( \mu_{\rm P} \) remains given by Eq.~\eqref{eq:muP_vp1}, but the probe temperature must be determined self-consistently as~\cite{meair2014local}
\begin{align}
	T_{\rm P}^{\rm VTP} &= -\frac{1}{\kappa_{\rm PP}} \sum_{\alpha \neq P} \left[ \Ltilde{1,1}_{{\rm P}\alpha} \mu_\alpha + \Ltilde{2,1}_{{\rm P}\alpha} \frac{T_\alpha}{T_0} \right],
\end{align}
where the effective thermal conductance is defined as
\begin{equation}
	\kappa_{\rm PP} = \frac{1}{T_0} \left( \mathcal{L}^{(2)}_{\rm PP} - \frac{[\mathcal{L}^{(1)}_{\rm PP}]^2}{\mathcal{L}^{(0)}_{\rm PP}} \right) = \frac{1}{T_0} \Ltilde{2,1}_{\rm PP}.
\end{equation}
With both \( \mu_{\rm P} \) and \( T_{\rm P}^{\rm VTP} \) determined, the terminal current takes the form
\begin{equation}
	I_{\rm R, VTP}^{(\nu)} = A^{(\nu)} (\mu_{\rm L} - \mu_{\rm R}) + \frac{1}{T_0} B^{(\nu)} (T_{\rm L} - T_{\rm R}),
	\label{eq:IRVTP}
\end{equation}
where the probe's influence is encoded in the effective coefficients:
\begin{align}
	A^{(\nu)} &= \Ltilde{\nu,\nu}_{\rm RL} + \frac{ \Ltilde{\nu+1,1}_{\rm RP} \Ltilde{1,1}_{\rm PL} }{ -\Ltilde{2,1}_{\rm PP} }, \nonumber \\
	B^{(\nu)}&= \Ltilde{\nu+1,\nu}_{\rm RL} + \frac{ \Ltilde{\nu+1,\nu}_{\rm RP} \Ltilde{2,1}_{\rm PL} }{ -\Ltilde{2,1}_{\rm PP} }.
\end{align}

The essential difference between VP and VTP models lies in their thermodynamic consistency. VPs enforce local charge neutrality by adjusting the probe potential to ensure \( I_{\rm P}^{(0)} = 0 \), but they do not enforce local thermal equilibrium; the probe temperature \( T_{\rm P}^{\rm VP} \) remains fixed, typically at the bath value, even when a net heat current \( I_{\rm P}^{(1)} \neq 0 \) flows. This mismatch can significantly affect predicted quantities such as the local temperature, \cite{bergfield2013probing, shastry2020scanning,shastryColdSpotsQuantum2015} electrochemical potential, \cite{bergfield2014thermoelectric} and local entropy. \cite{shastry2016temperature, Stafford2016local,shastryColdSpotsQuantum2015}

For example, the chemical potential difference between the VP and VTP models is given by\cite{bergfield2014thermoelectric}
\begin{equation}
	\mu_{\rm P}^{\mathrm{VP}} - \mu_{\rm P}^{\mathrm{VTP}} = \left( T_{\rm P}^{\mathrm{VP}} - T_{\rm P}^{\mathrm{VTP}} \right) \cdot \left(-\frac{1}{T_0} \frac{\mathcal{L}^{(1)}_{\rm PP}}{\mathcal{L}^{(0)}_{\rm PP}} \right),
	\label{eq:mu_difference}
\end{equation}
where the right-hand factor reflects the strength of thermoelectric coupling at the probe. In systems where \( \mathcal{L}^{(1)}_{\rm PP} \) is appreciable, even moderate differences in local temperature can induce substantial errors in the predicted probe potential, highlighting the importance of energy conservation in dephasing models.\cite{bergfield2014thermoelectric, bergfield2013probing}

Similarly, the difference in effective two-terminal current caused by probe choice is given by
\begin{equation}
	\delta I_R^{(\nu)} = I_{\rm R, VP}^{(\nu)} - I_{\rm R, VTP}^{(\nu)} = \frac{ \widetilde{\mathcal{L}}^{(\nu+1,1)}_{\rm PR} }{ \widetilde{\mathcal{L}}^{(2,1)}_{\rm PP} } I_{\rm P,VP}^{(1)},
	\label{eq:deltaIR_electronic}
\end{equation}
where
\begin{align}
	I_{\rm P,VP}^{(1)} &= \Ltilde{1,1}_{\rm PL}\mu_L +\Ltilde{1,1}_{\rm PR} \mu_R \nonumber \\
 &+\frac{1}{T_0} \left[ \widetilde{\mathcal{L}}^{(2,1)}_{\rm PL} (T_L - T_{\rm P}^{\rm VP}) + \widetilde{\mathcal{L}}^{(2,1)}_{\rm PR} (T_R - T_{\rm P}^{\rm VP}) \right]
	\label{eq:probe_heatcurrent}
\end{align}
is the net heat current flowing into the VP. Although this form involves individual potentials, the resulting expressions remain gauge invariant by virtue of the sum rule in Eq.\ref{eq:effOnsager_gauge_invariance}. The correction \( \delta I_R^{(\nu)} \) is therefore governed by two quantities: the probe heat current, and the ratio of generalized Onsager coefficients. 

The heat current \( I_{\rm P,VP}^{(1)} \) reflects the energetic imbalance introduced by externally fixing the probe temperature. Under broken thermal or structural symmetry, this imbalance can become substantial: the VP cannot locally equilibrate, instead acting as a spurious source or sink of energy. In such cases, it violates the high-impedance condition expected of a probe and perturbs transport in the other terminals. In extreme scenarios, the VP fails to model decoherence altogether, as it cannot suppress interference features despite strong coupling (see Appendix). By contrast, the VTP adjusts both its temperature and chemical potential self-consistently, ensuring local equilibrium and minimizing entropy production.

The degree to which the VP’s imbalance distorts transport depends on the effective Onsager ratio \( \widetilde{\mathcal{L}}^{(\nu+1,1)}_{\rm PR} / \widetilde{\mathcal{L}}^{(2,1)}_{\rm PP} \), which governs the coupling between probe heat flow and current in terminal \( R \). This ratio is maximized when the probe exhibits strong nonlocal thermoelectric coupling but weak local thermal conductance, i.e. the regime where even moderate \( I_{\rm P,VP}^{(1)} \) can induce disproportionate nonlocal effects. Such amplification is largest near interference nodes or under asymmetric thermal bias, where local thermodynamic consistency becomes essential.

These effects collectively explain the divergence between VP and VTP predictions. While charge current discrepancies are often suppressed due to particle conservation, heat current errors can grow dramatically, revealing a fundamental thermodynamic inconsistency. In asymmetric setups, such deviations can span several orders of magnitude, highlighting the VP’s failure to simulate decoherence in physically meaningful terms.

\subsection{Pure Thermal Circuits}

Next, we consider the complementary case of a purely thermal circuit, in which charge currents vanish at each terminal. Specifically, the electrochemical potentials \( \mu_{\rm L} \) and \( \mu_{\rm R} \) are dynamically adjusted such that \( I^{(0)}_\alpha = 0 \) for all \( \alpha \). Imposing this open-circuit condition reduces the linear-response problem to solving the matrix equation \( \mathbf{A} \vec{x} = \mathbf{B} \), where
\begin{equation}
	\mathbf{A}
	= \begin{pmatrix}
		\mathcal{L}^{(0)}_{\rm RR} & \mathcal{L}^{(0)}_{\rm RP} \\
		\mathcal{L}^{(0)}_{\rm PR} & \mathcal{L}^{(0)}_{\rm PP}
	\end{pmatrix}, 
	\mathbf{B}_\alpha = -\frac{1}{T_0} \sum_\beta \mathcal{L}^{(1)}_{\alpha\beta} (T_\beta - T_\alpha),
\end{equation}
and $\vec{x} = (\mu_{\rm R}, \mu_{\rm P})$ with \( \alpha = {\rm R, P} \). Without loss of generality, we fix the gauge by setting \( \mu_{\rm L} = 0 \).
%
%
Solving this system yields the chemical potentials at the right terminal and probe
\begin{align}
	\mu_{\rm R} &= -\frac{1}{\Ltilde{0,0}_{\rm RR}} \left[ \Ltilde{1,0}_{\rm RL} \frac{T_{\rm L} - T_{\rm R}}{T_0} + \Ltilde{1,0}_{\rm RP} \frac{T_{\rm P} - T_{\rm R}}{T_0} \right], \\
	\mu_{\rm P} &= \frac{1}{\Lfn{0}_{\rm PP} \Ltilde{0,0}_{\rm RR}} \left(\Lfn{0}_{\rm RR} \mathbf{B}_{\rm P} - \Lfn{0}_{\rm PR}\mathbf{B}_{\rm R} \right)
%
\end{align}
where the $\det \mathbf{A}$ has been expressed as $\Lfn{0}_{\rm RR} \Lfn{0}_{\rm PP} - (\Lfn{0}_{\rm RP})^2 \equiv \Lfn{0}_{\rm PP} \Ltilde{0,0}_{\rm RR}$.

Regardless of probe type, the heat current into terminal $\alpha$ then takes the familiar form
\begin{equation}
	I_\alpha^{(1)} = \sum_\beta \tilde{\kappa}_{\alpha \beta} (T_\beta - T_\alpha),
	\label{eq:purethermal_heatcurrent}
\end{equation}
where the effective thermal conductance tensor $\tilde{\kappa}_{\alpha\beta}$, evaluated under the open-circuit condition, is given by\cite{bergfield2013probing}
\begin{align}
	&\tilde{\kappa}_{\alpha\beta} = \frac{1}{T_0} \left[ \Lfn{2}_{\alpha\beta}
	- \frac{(\Lfn{1}_{\alpha\beta})^2}{ \Ltilde{0,0}_{\alpha\beta}} \right. \nonumber \\
 & \; \;- \left. \Lfn{0}
	\left(
	\frac{\Lfn{1}_{\alpha\gamma} \Lfn{1}_{\alpha\beta}}{\Lfn{0}_{\alpha\gamma} \Lfn{0}_{\alpha\beta}} +
	\frac{\Lfn{1}_{\gamma\beta} \Lfn{1}_{\alpha\beta}}{\Lfn{0}_{\gamma\beta} \Lfn{0}_{\alpha\beta}} -
	\frac{\Lfn{1}_{\alpha\gamma} \Lfn{1}_{\gamma\beta}}{\Lfn{0}_{\alpha\gamma} \Lfn{0}_{\gamma\beta}}
	\right)
	\right],
	\label{eq:kappatilde}
\end{align}
with $\alpha$, $\beta$, and $\gamma$ labeling distinct terminals. For convenience, we define the reciprocal series combination
\begin{equation}
	\frac{1}{\Lfn{0}} = \frac{1}{\Lfn{0}_{12}} + \frac{1}{\Lfn{0}_{13}} + \frac{1}{\Lfn{0}_{23}}.
	\label{eq:L0_series}
\end{equation}

As in the electronic case, transport in a purely thermal circuit is highly sensitive to the boundary conditions imposed at the probe. In the VP configuration, the probe temperature \( T_{\mathrm{P}}^{\mathrm{VP}} \) is externally specified. By contrast, in the VTP configuration, the probe temperature is determined self-consistently by enforcing both charge and heat current conservation. This yields\cite{bergfield2013probing}
\begin{equation}
	T_{\mathrm{P}}^{\mathrm{VTP}} = \frac{\tilde{\kappa}_{\mathrm{PL}} T_{\mathrm{L}} + \tilde{\kappa}_{\mathrm{PR}} T_{\mathrm{R}}}{\tilde{\kappa}_{\mathrm{PL}} + \tilde{\kappa}_{\mathrm{PR}}}.
\end{equation}
%
Plugging these terms into Eq.~\ref{eq:purethermal_heatcurrent}, the heat current flowing into the right terminal for a VP 
\begin{equation}
	I_{\rm R,VP}^{(1)} = \tilde{\kappa}_{\rm RL} (T_{\rm L} - T_{\rm R}) + \tilde{\kappa}_{\rm RP} (T_{\rm P}^{\rm VP} - T_{\rm R}),
\end{equation}
and for a VTP
\begin{equation}
	I_{\rm R,VTP}^{(1)} = \left( \tilde{\kappa}_{\rm RL} + \frac{\tilde{\kappa}_{\rm RP} \tilde{\kappa}_{\rm PL}}{\tilde{\kappa}_{\rm PL} + \tilde{\kappa}_{\rm RP}} \right) (T_{\rm L} - T_{\rm R}),
\end{equation}
which reflects the expected effective two-terminal form, with the VTP mediating the additional thermal pathway.

The difference in heat current between VP and VTP configurations is given by 
\begin{align}
	\delta I_{\rm R}^{(1)} &= I_{\rm R,VP}^{(1)} - I_{\rm R,VTP}^{(1)} = \tilde{\kappa}_{\rm RP} (T_{\rm P}^{\rm VP} - T_{\rm P}^{\rm VTP}) \nonumber \\
	&= - \frac{\tilde{\kappa}_{\rm RP}}{\tilde{\kappa}_{\rm PL} + \tilde{\kappa}_{\rm RP}} I^{(1)}_{\rm P,VP}
\end{align}
where the VP heat current is given by
\begin{equation}
	I^{(1)}_{\rm P,VP} = \tilde{\kappa}_{\rm PR}(T_{\rm R}-T_{\rm P}^{\rm VP}) + \tilde{\kappa}_{\rm PL}(T_{\rm L} - T_{\rm P}^{\rm VP}).
\end{equation}
This vanishes only if the thermal currents from the left and right leads cancel exactly, a condition rarely satisfied in realistic systems. As in the electronic circuit case, the magnitude of this correction is enhanced by asymmetry in either tunnel-coupling or thermal bias.

\begin{figure}[!htb]
	\centering
	\begin{overpic}[width=.9\linewidth]{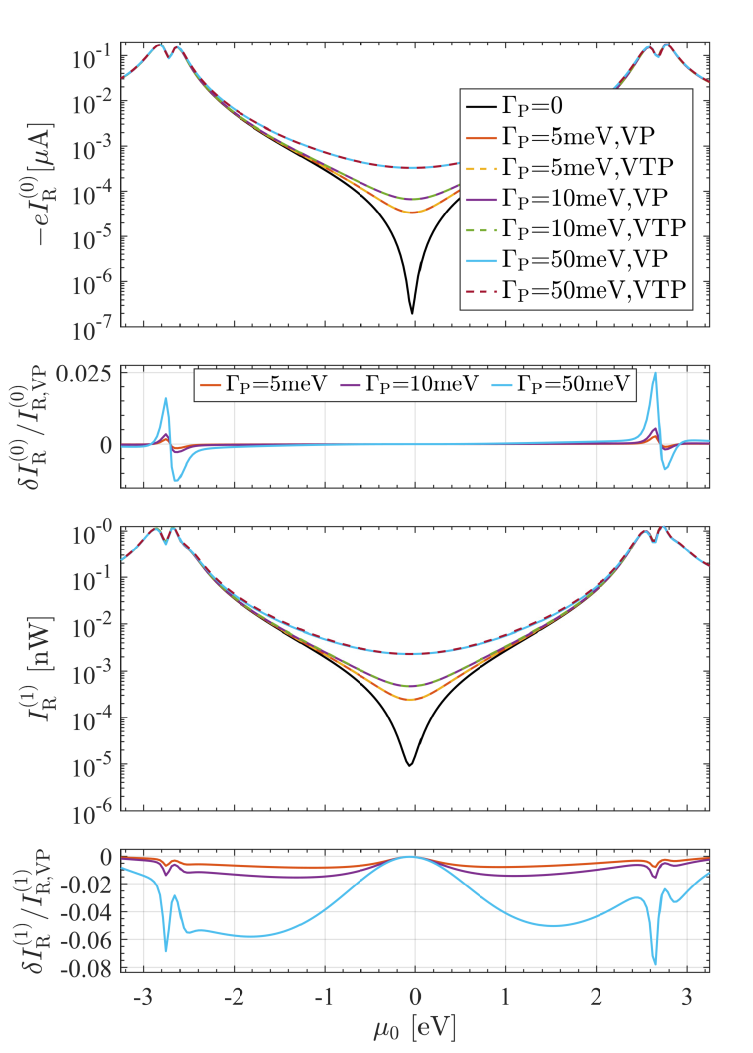} 
		\put(12,69){\includegraphics[width=1in]{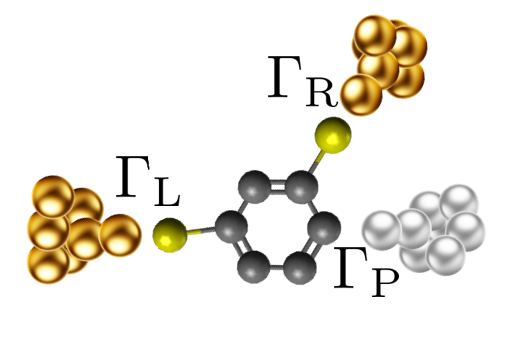}}
	\end{overpic}
	\caption{
		Calculated charge and heat currents through a three-terminal (1,3)-BDT junction under symmetric thermal bias ($T_{\rm L}$=325K, $T_{\rm R}$=275K) and symmetric lead–molecule coupling ($\Gamma_{\rm L}$=$\Gamma_{\rm R}$= 0.5eV), with a fixed electrical bias ($\mu_{\rm L}$+5meV, $\mu_{\rm R}$=-5meV). Decoherence is introduced via a local probe acting as either a VP or VTP, with coupling strength \( \Gamma_{\rm P} \).
		In this symmetric configuration, both models lift the interference node at \( \mu_0 = 0 \) in a nearly identical manner, yielding charge and heat currents that agree to within $\sim$2.5\% and $\sim$7.8\%, respectively, at $\Gamma_{\rm P}$=50meV. 
		%
However, this apparent agreement is {\em accidental}, arising only from the specific spatial and thermal symmetry of this setup, which masks the fundamental thermodynamic differences between the two models, as demonstrated in subsequent figures.
			}
				\label{fig:fig1_sym}
\end{figure}

This deviation also appears in observable quantities such as the thermopower. In our gauge, the Seebeck coefficient is defined as 
\begin{equation}
	S = \frac{1}{e T_0} \frac{\mu_{\mathrm{R}}}{\Delta T},
\end{equation}
so the model-dependent correction is
\begin{align}
	\delta S &= \frac{1}{e T_0 \Delta T} \left( \mu_{\mathrm{R}}^{\mathrm{VP}} - \mu_{\mathrm{R}}^{\mathrm{VTP}} \right) \nonumber \\
	&= \frac{1}{e T_0 \Delta T} \cdot \frac{ \widetilde{\mathcal{L}}^{(1,0)}_{\mathrm{RP}} }{ \widetilde{\mathcal{L}}^{(0,0)}_{\mathrm{RR}} } \cdot \left( T_{\mathrm{P}}^{\mathrm{VTP}} - T_{\mathrm{P}}^{\mathrm{VP}} \right).
\end{align}
Although the internal probe temperature \( T_{\mathrm{P}} \) can differ significantly between VP and VTP models, symmetry constraints typically suppress the corresponding correction to \( S \), particularly away from resonance. As an energy-weighted average evaluated under the condition \( I^{(0)}_{\mathrm{R}} = 0 \), the thermopower is less sensitive to probe-induced perturbations than either the raw heat currents or the entropy production.


\begin{figure*}[!htb]
	\subfloat[Asymmetric thermal bias; $T_{\rm L}$=350K, $T_{\rm R}$=300K, $T_0$=300K]{\includegraphics[width=.4\linewidth]{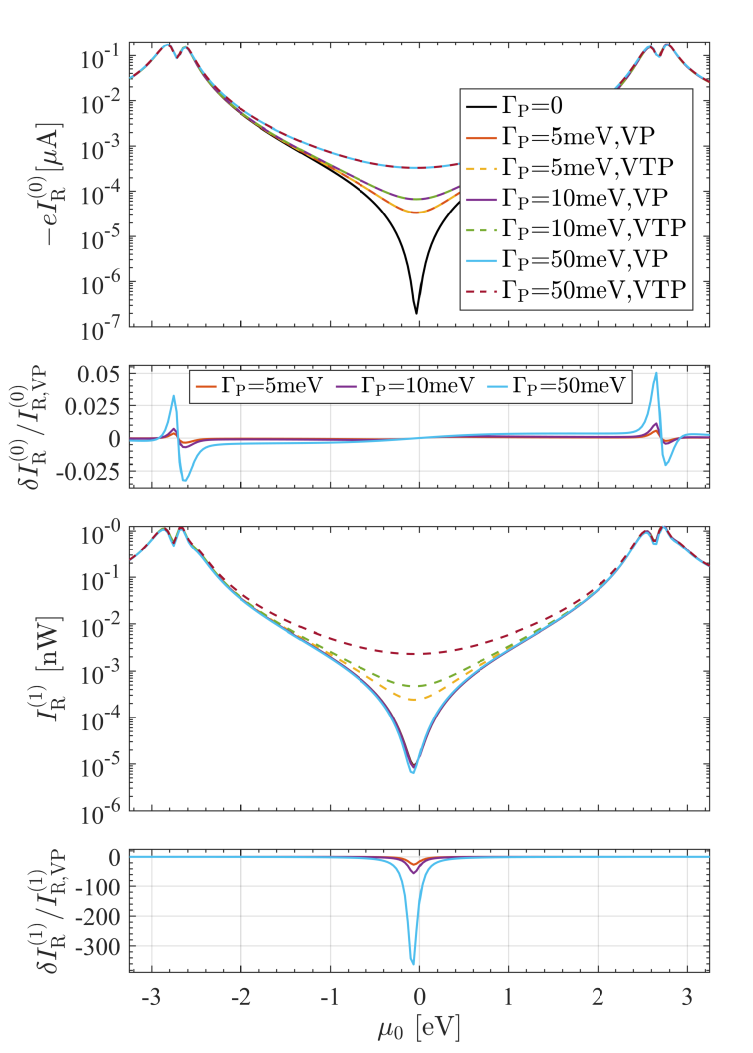}} 
 \subfloat[Asymmetric tunnel coupling; $\Gamma_{\rm L}$= 0.5eV, $\Gamma_{\rm R}$ = 0.25eV]{\includegraphics[width=.4\linewidth]{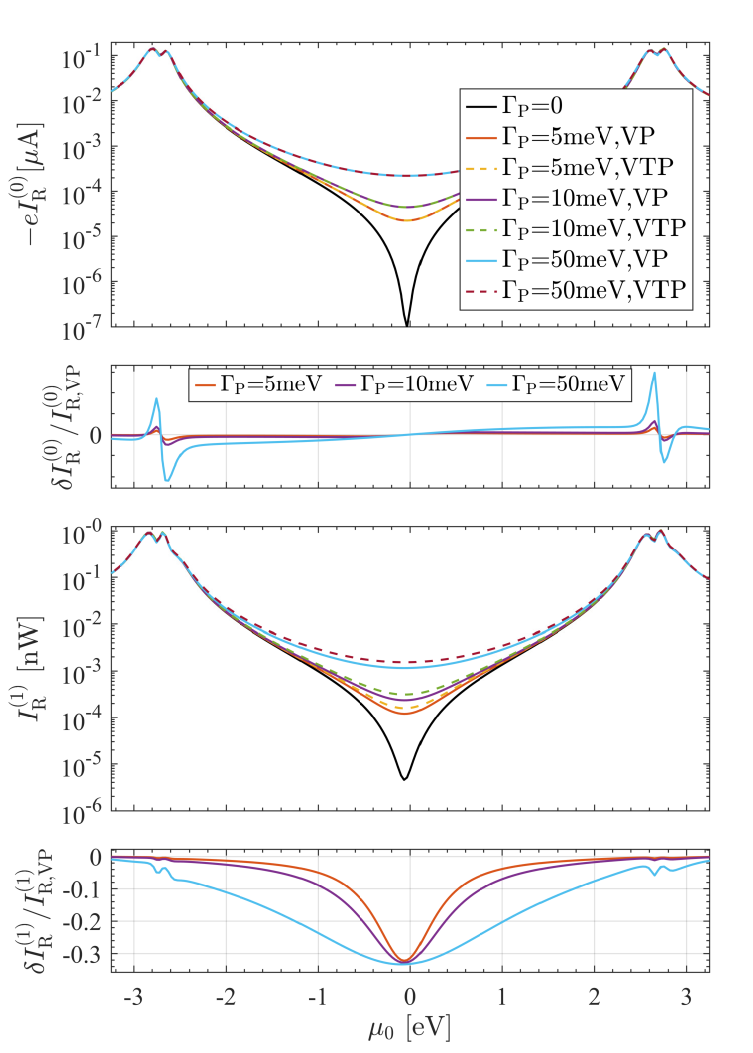}} 
\caption{
	Calculated charge and heat currents through a 1,3-BDT junction are shown for two representative sources of asymmetry and a range of probe couplings \( \Gamma_{\rm P} \), using both VP and VTP dephasing models. In all cases, a fixed electrical bias is applied ($\mu_{\rm L}$ = +5meV, $\mu_{\rm R}$=-5meV).
	\textbf{(a)} Under asymmetric thermal bias, both probes lift the node's charge current effect similarly, with peak relative errors of just 5.1\% near resonance. In stark contrast, only the VTP lifts the heat current node; the VP fails entirely and \emph{does not act as a dephasing probe}, even at large \( \Gamma_{\rm P} \), with the node persisting and relative errors exceeding {36,204\%}.
	{\bf (b)} Under asymmetric tunnel couplings, both probes lift the node, but notable discrepancies emerge: heat currents differ by up to 33.3\%, while charge current errors remain below 1.5\%. These further indicate that agreement between VP and VTP models arises only under special symmetry conditions and does not reflect a general equivalence.
}
	\label{fig:fig2_asym}
\end{figure*}

These differences highlight a deeper thermodynamic inequivalence between the probe models. According to Eq.~\ref{eq:entropy_production}, the excess entropy produced in the VP configuration relative to the VTP is
\begin{equation}
	\dot{S}_{\mathrm{VP}} - \dot{S}_{\mathrm{VTP}} = \frac{ I^{(1)}_{\mathrm{P,VP}} }{ T_{\mathrm{P}}^{\mathrm{VP}} }.
\end{equation}
Thus, the VP heat current term also
quantifies the irreversible entropy generation arising from net energy exchange between the system and a probe held at fixed temperature. Though this contribution may be numerically small in symmetric or weakly biased regimes, its mere presence reveals a conceptual inconsistency: the VP functions not as a passive, locally equilibrated dephasing bath, but rather as an active thermal reservoir capable of injecting or extracting entropy.

From a thermodynamic perspective, this is a crucial distinction. While both models may suppress phase coherence, only the VTP does so without violating local equilibrium constraints. It enforces zero net exchange of both charge and energy, ensuring that any entropy production originates solely from the imposed thermodynamic bias between physical leads. The VP, by contrast, permits entropy flow with an artificial reservoir whose temperature is not determined by system dynamics, an assumption that can distort both physical interpretation and observable response.


It is important to emphasize, however, that the absence of entropy exchange with the VTP does not imply that dephasing is thermodynamically free. Phase randomization corresponds to a loss of quantum information, which in a microscopic model would necessarily entail entropy generation through coupling to additional degrees of freedom. The VTP abstracts away these hidden variables, offering a ``coarse-grained'' description that preserves transport-level thermodynamic structure while concealing the underlying irreversibility.

\section{Example System: The Au–1,3-benzenedithiol–Au Junction}

To illustrate the thermodynamic impact of different dephasing models, we calculate charge and energy transport in the archetypal Au–1,3-benzenedithiol–Au (1,3-BDT) molecular junction configured with a third probe terminal. In the geometry considered, the locally coupled left (L) and right (R) electrodes are meta-configured while the VP or VTP probe are para-configured to L and ortho-configured to R, as indicated in Fig.~\ref{fig:fig1_sym}.
This configuration is known to exhibit a pronounced mid-gap transmission node at \( \mu_0 = 0 \), originating from destructive interference between symmetry-related $\pi$-orbitals in the conjugated backbone which can be lifted via dephasing.\cite{cardamone2006controlling,stafford2007quantum} 
Transport is computed within the H\"uckel+NEGF framework introduced above, in the linear-response regime with the VP probe temperature set to ambient temperature ($T_0 = 300$~K). 

While the results presented here focus on a locally coupled probes, spatially extended probe couplings yield qualitatively similar behavior (See Appendix), reinforcing the generality of these conclusions.
%
In the sections that follow, we systematically explore how asymmetries in thermal bias, tunnel coupling, and probe strength modulate the response of the junction, and how the underlying thermodynamic assumptions of the VP and VTP lead to divergent predictions, both qualitatively and quantitatively, when quantum coherence plays a central role.

%
%

\begin{figure*}[!tb]
\subfloat[Influence of Thermal Bias Symmetry]{\includegraphics[width=.42\linewidth]{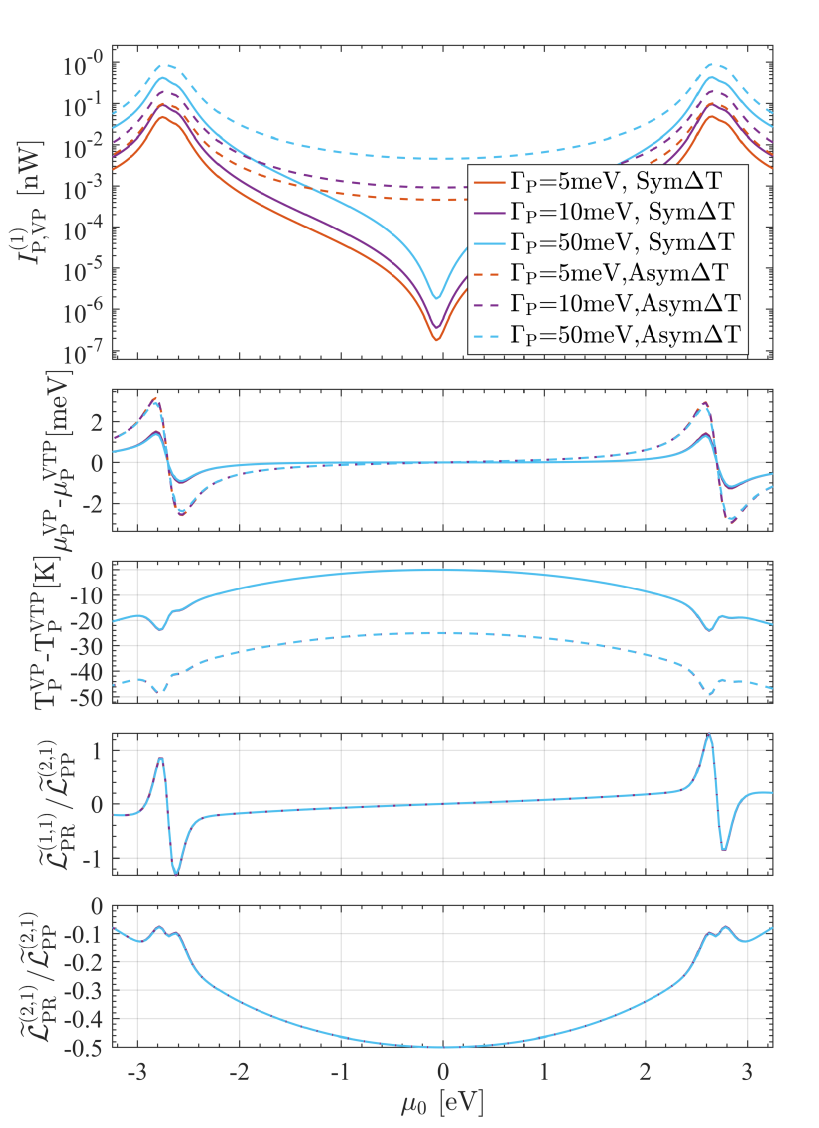}}
\subfloat[Influence of Tunnel-coupling Symmetry]{\includegraphics[width=.42\linewidth]{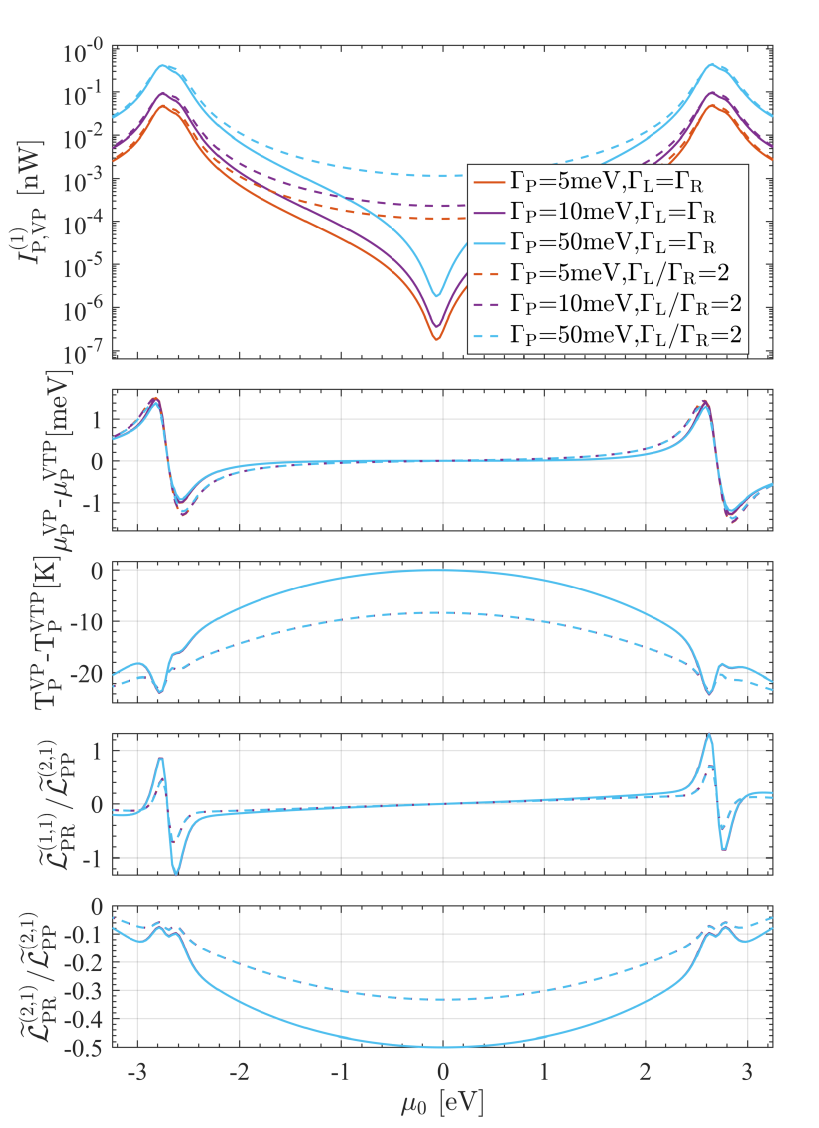}}
\caption{
	Contributions to the current correction \( \delta I_R^{(\nu)} \) from Eq.~\ref{eq:deltaIR_electronic} for the 1,3-BDT junction, including VP heat current \( I_{\rm P,VP}^{(1)} \), deviations of \( \mu_{\rm P} \) and \( T_{\rm P} \) between the two models, and the effective Onsager ratio, are plotted versus level alignment \( \mu_0 \) for various probe couplings \( \Gamma_{\rm P} \). In all cases, a fixed electrical bias is applied ($\mu_{\rm L}$ = +5meV, $\mu_{\rm R}$=-5meV).
	{\bf (a)} Thermal asymmetry induces large deviations in \( \mu_{\rm P} \) and \( T_{\rm P} \), while leaving the Onsager ratio mostly unchanged.	{\bf (b)} Asymmetric tunnel coupling reduces deviations in the Onsager ratios (and deviations in \( T_{\rm P} \), but minimally affects \( \mu_{\rm P} \).
	In both cases, symmetry breaking removes the nodal suppression of \( I_{\rm P,VP}^{(1)} \) at \( \mu_0 = 0 \), allowing the probe to exchange energy with the system.
	Variations in the Onsager ratios closely match deviations in the probe's thermodynamic variables, highlighting their role in converting thermodynamic inconsistencies into current corrections.	
}

%
%
%

\label{fig:fig3_terms_that_influence_asym}
\end{figure*}

\begin{figure*}[tb]
\subfloat[Symmetric Junction \newline($\Gamma_{\rm L}$=$\Gamma_{\rm R}$=0.5eV; $T_{\rm L}$=325, $T_{\rm R}$=275K)]{\includegraphics[width=.32\linewidth]{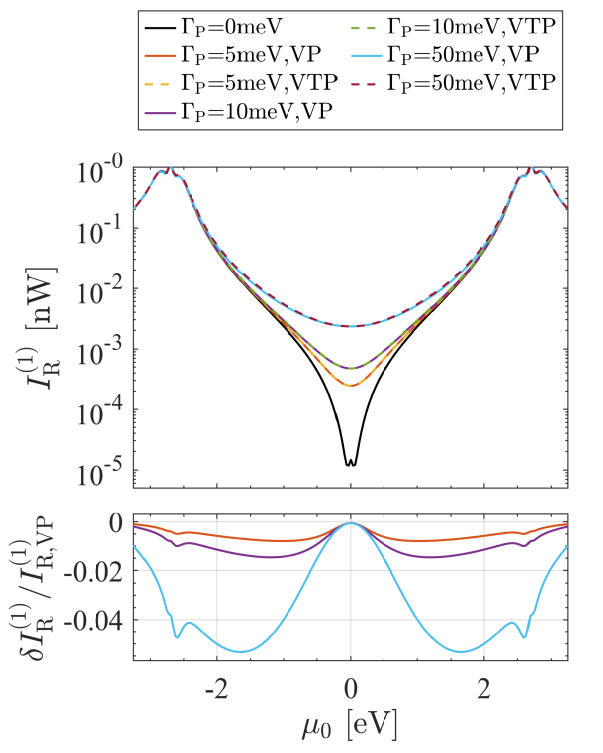}}
\subfloat[Asymmetric Thermal Bias \newline ($\Gamma_{\rm L}$=$\Gamma_{\rm R}$=0.5eV; $T_{\rm L}$=350K, $T_{\rm R}$=275K)]{\includegraphics[width=.32\linewidth]{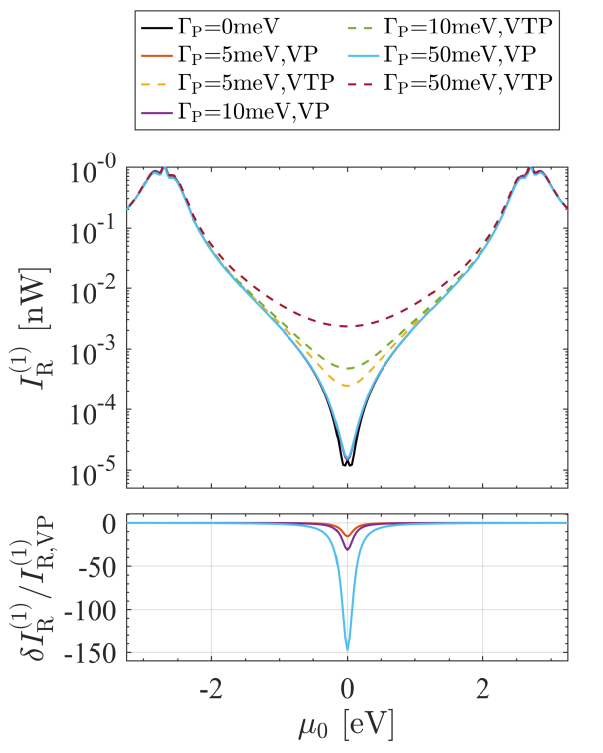}}
\subfloat[Asymmetric Tunnel-Coupling \newline ($\Gamma_{\rm L}$=0.5eV,$\Gamma_{\rm R}$=0.25eV; $T_{\rm L}$=325K, $T_{\rm R}$=275K)]{\includegraphics[width=.32\linewidth]{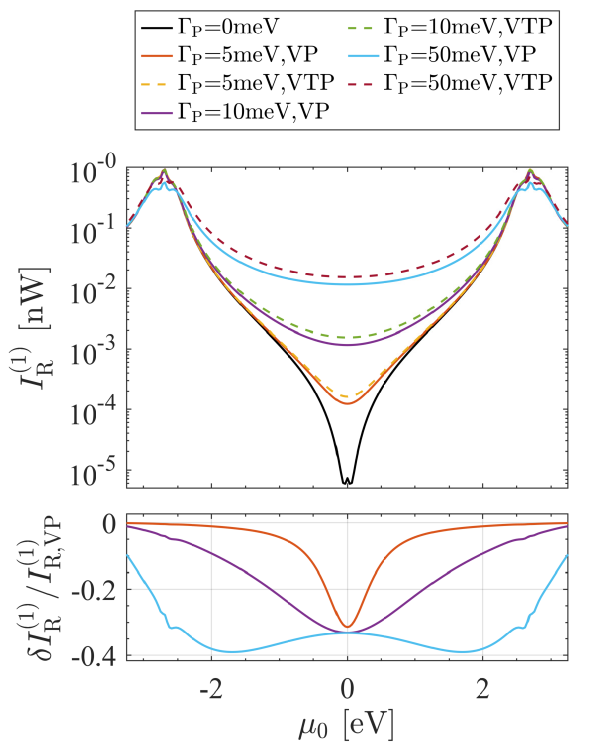}}
\caption{
		Heat current into the right-hand electrode of a 1,3-BDT junction under pure-thermal (i.e. open electrical) circuit conditions, computed using VP and VTP models for a range of probe-molecule coupling strengths $\Gamma_{\rm P}$ across three configurations: 
		{\bf (a)} symmetric temperature and symmetric tunnel coupling, 
		{\bf (b)} asymmetric temperature and symmetric tunnel coupling, and 
		{\bf (c)} symmetric temperature and asymmetric tunnel coupling. 
		Relative errors between VP and VTP predictions remain modest in panel (a), peaking at $\sim$4.5\%, but grow drastically in panel (b) to $\sim$14,723\%, reflecting a fundamental breakdown of the VP model under asymmetric thermal bias. 
		Coupling asymmetry in panel (c) produces intermediate discrepancies, with heat current errors reaching $\sim$40\%. 
		Notably, in panel (b), the VP fails to lift the interference node in the heat current with increasing \( \Gamma_{\rm P} \), underscoring its inability to model local equilibration under broken symmetry. 
	}
	\label{fig:purethermaliq}
	\end{figure*}

\subsection{Electronic Circuit}
%
We begin by examining the electronic response of the three-terminal 1,3-BDT junction under symmetric tunneling (\( \Gamma_{\rm L} = \Gamma_{\rm R} = 0.5~\mathrm{eV} \)) and thermal bias (\( T_{\rm L} = 325~\mathrm{K}, T_{\rm R} = 275~\mathrm{K} \)), with a fixed voltage bias (\( \mu_{\rm L} = +5~\mathrm{meV}, \mu_{\rm R} = -5~\mathrm{meV} \)). Figure~\ref{fig:fig1_sym} shows the resulting charge and heat currents into the right electrode.
In the fully coherent limit (\( \Gamma_{\rm P} = 0 \)), destructive interference at \( \mu_0 = 0 \) suppresses both charge and heat transport, producing a pronounced dip near the node. Introducing dephasing via finite probe coupling \( \Gamma_{\rm P} \) lifts the node as expected. In this symmetric configuration, the VP and VTP models yield similar predictions: charge current discrepancies remain below 2.5\%, and heat current deviations stay below 8\% 
across the spectrum.

While this agreement might suggest that the distinction between VP and VTP models is subtle, this is not the case. The apparent similarity arises only under conditions of exact symmetry, where the thermal inflows from the left and right leads nearly cancel. With the VP temperature fixed at the midpoint $T_0$, the net heat current into the probe vanishes by construction, enabling it to emulate the VTP despite violating local thermodynamic equilibrium. This agreement is accidental, not fundamental.

This fragile equivalence collapses when thermal symmetry is broken. Figure~\ref{fig:fig2_asym}(a) shows the result of applying an asymmetric temperature bias (\( T_{\rm L} = 350~\mathrm{K}, T_{\rm R} = 300~\mathrm{K} \)). While the VTP correctly lifts the interference node and restores energy flow as decoherence increases, the VP fails catastrophically: the heat current remains suppressed near \( \mu_0 = 0 \), even as $\Gamma_{\rm P}$ is increased. In the case where $T_{\rm R} = T_{\rm P}^{\rm VP}=T_0$, the node's effect on the heat current into the right lead is entirely independent of $\Gamma_{\rm P}$ (See Appendix). Relative errors of $\delta I_{\rm R}^{(1)}$ between VTP and VP models exceed 36,204\%, signaling not just a quantitative discrepancy but a thermodynamic breakdown. Constrained to a fixed temperature, the VP cannot equilibrate with its local environment, violating energy conservation and acting as an unphysical entropy sink. In this regime, it does not simulate decoherence, it suppresses it. 

A similar breakdown occurs with coupling asymmetry. Figure~\ref{fig:fig2_asym}(b) shows results for asymmetric tunnel couplings (\( \Gamma_{\rm L} = 0.5~\mathrm{eV}, \Gamma_{\rm R} = 0.25~\mathrm{eV} \)) with symmetric thermal bias and the same voltage bias as before. 
%
At the nodal energy, the transmission from L to R terminal is strongly suppressed, and heat injected by the hot electrode flows primarily into the probe such that $ I_{\rm P,VP}^{(1)} \approx I_{\rm R,VP}^{(1)}$. Then, in accordance with Eq.~\ref{eq:deltaIR_electronic}, this results in a relative heat current correction given by \( \widetilde{\mathcal{L}}^{(2,1)}_{\rm PR} / \widetilde{\mathcal{L}}^{(2,1)}_{\rm PP} \) which, near the node, simplifies via Sommerfeld expansion to
\begin{equation}
	\frac{\delta I_{\rm R}^{(1)}}{I_{\rm R, VP}^{(1)}} \approx - \frac{\Gamma_{\rm L} - \Gamma_{\rm R}}{\Gamma_{\rm L} + \Gamma_{\rm R}}
\end{equation}
as derived in the Appendix. In the configuration considered here this gives the observed -33.3\% peak relative error which, interestingly, is also independent of $\Gamma_{\rm P}$. This error vanishes in the case of symmetric coupling due to a precise balance of heat currents which cancel the heat current flowing into the probe.

To further clarify the origin of the discrepancies between VP and VTP models, we examine the key quantities entering Eq.~\eqref{eq:deltaIR_electronic} in Fig.~\ref{fig:fig3_terms_that_influence_asym} under the aforementioned asymmetries. Specifically, we plot the VP heat current \( I_{\rm P,VP}^{(1)} \), deviations in the probe's chemical potential and temperature (\( \mu_{\rm P}^{\rm VP} - \mu_{\rm P}^{\rm VTP} \), \( T_{\rm P}^{\rm VP} - T_{\rm P}^{\rm VTP} \)), and the effective Onsager ratios $\Ltilde{\nu+1,1}_{\rm PR} / \Ltilde{2,1}_{\rm PP}$, $\nu=0,1$, all as functions of level alignment \( \mu_0 \) for various probe couplings \( \Gamma_{\rm P} \).
In panel (a), thermal asymmetry induces large deviations in both \( \mu_{\rm P} \) and \( T_{\rm P} \), while the Onsager ratios are unaffected. Conversely, in panel (b), tunnel-coupling asymmetry leads to marked changes in the Onsager ratio and \( T_{\rm P} \), but has little impact on \( \mu_{\rm P} \). 

In both cases, broken symmetry lifts the nodal suppression of \( I_{\rm P,VP}^{(1)} \) near \( \mu_0 = 0 \). As indicated by Eq.\ref{eq:deltaIR_electronic}, it is precisely this suppression which masks the transport signatures of the thermodynamic discrepancies between VP and VTP models. 
%
Once the symmetry is broken, however, the exact balance of of heat transport is lost and the inequivalence between the two models become manifest.


\begin{figure*}[!tb]
\subfloat[Symmetric Junction \newline ($\Gamma_{\rm L}$=$\Gamma_{\rm R}$=0.5eV; $T_{\rm L}$=325, $T_{\rm R}$=275K)]{\includegraphics[width=.33\linewidth]{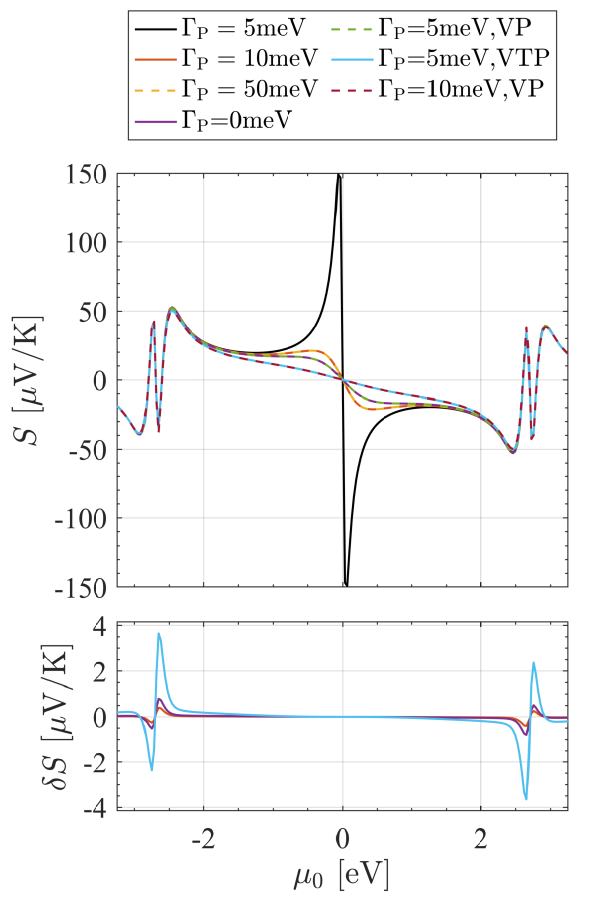}}
\subfloat[Asymmetric Thermal Bias \newline \centering{($\Gamma_{\rm L}$=$\Gamma_{\rm R}$=0.5eV; $T_{\rm L}$=350K, $T_{\rm R}$=275K)}]{\includegraphics[width=.33\linewidth]{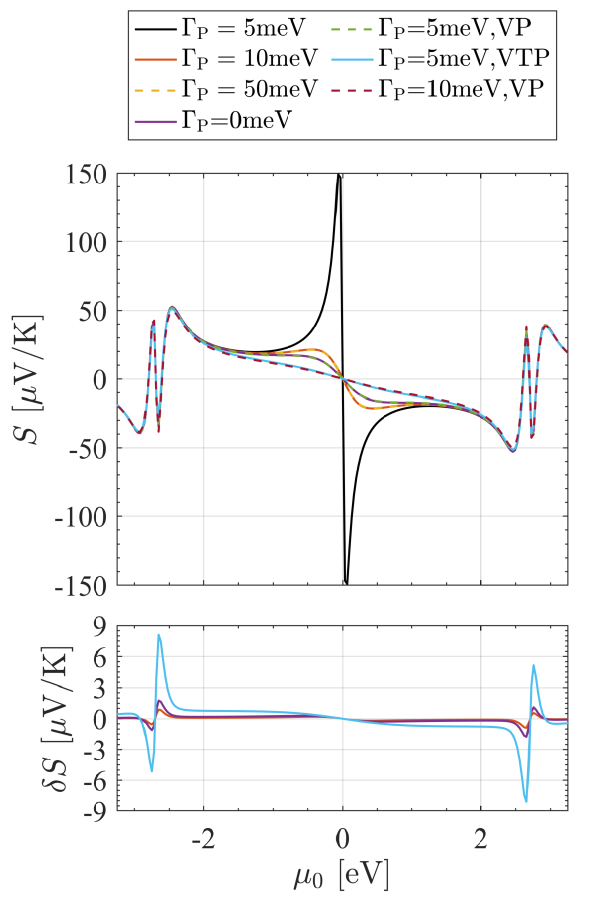}}
\subfloat[Asymmetric Tunnel-Coupling \newline ($\Gamma_{\rm L}$=0.5eV,$\Gamma_{\rm R}$=0.25eV; $T_{\rm L}$=325K, $T_{\rm R}$=275K)]{\includegraphics[width=.33\linewidth]{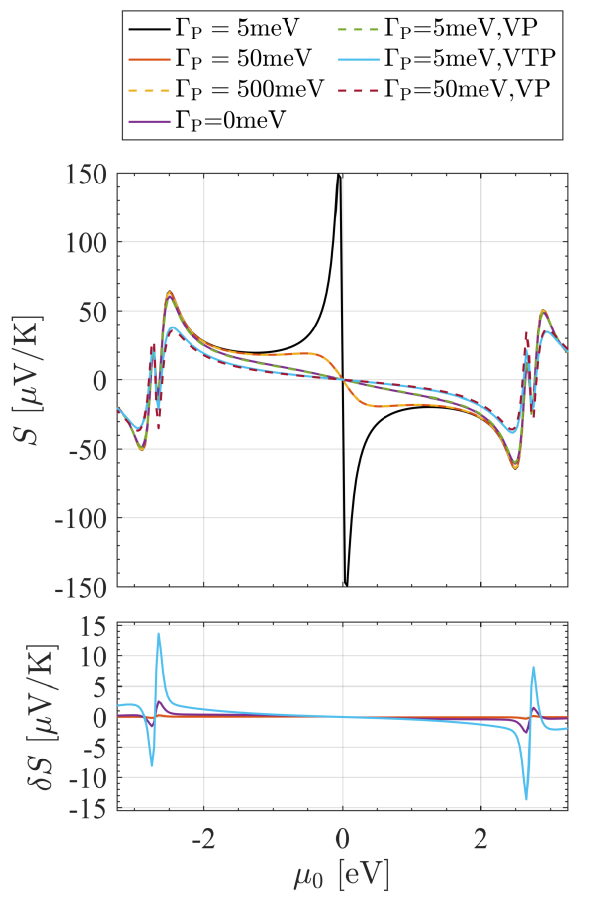}}
	\caption{
		Calculated Seebeck coefficient \( S \) between terminals L and R of a 1,3-BDT junction as a function of level alignment \( \mu_0 \), comparing VP and VTP probe models over a range of probe–molecule couplings \( \Gamma_{\rm P} \). 
		{\bf (a)} Under symmetric conditions, the models agree closely, with deviations below \( 3.6\,\mu\mathrm{V/K} \). 
		{\bf (b)} Asymmetric thermal bias increases discrepancies, with peak errors reaching \( 8.1\,\mu\mathrm{V/K} \). 
		{\bf (c)} Asymmetric tunnel coupling further amplifies the deviation, yielding a maximum difference of \( 13.6\,\mu\mathrm{V/K} \). 
		Despite significant discrepancies in heat current, the thermopower remains comparatively robust due to its open-circuit definition, which allows partial compensation of VP-induced inconsistencies.}
\label{fig:purethermal_thermopower}
\end{figure*}

\subsection{Pure Thermal Circuits}

We now turn to the case of a pure thermal circuit, i.e. a junction acting as an open electrical circuit. The heat current \( I^{(1)}_{\rm R} \) through a 1,3-BDT junction is shown as a function of level alignment \( \mu_0 \) and probe coupling \( \Gamma_{\rm P} \) in Fig.~\ref{fig:purethermaliq}, evaluated across three representative symmetry configurations.
In the fully symmetric case, shown in Fig.~\ref{fig:purethermaliq}(a), the VP and VTP yield nearly identical results, with heat current discrepancies remaining below 4.5\%. As in the electronic circuit, this agreement is not fundamental but instead is accidental: it arises from a cancellation of opposing thermal flows when the fixed VP temperature 
coincides with the midpoint of a symmetric thermal bias. Under these fine-tuned conditions, the VP appears to mimic local equilibrium despite lacking any mechanism to enforce it.

Once symmetry is broken, this equivalence collapses. Under asymmetric thermal bias, shown in Fig.~\ref{fig:purethermaliq}(b), the VTP lifts the interference node with increasing \( \Gamma_{\rm P} \), consistent with its role as a physically consistent dephasing mechanism. The VP, by contrast, fails qualitatively: the node persists even at strong coupling, and heat current errors exceed 14,723\%. This breakdown reflects the VP’s inability to exchange energy with its environment in a self-consistent manner, effectively converting it into an artificial entropy sink.

Asymmetric tunnel coupling, shown in Fig.~\ref{fig:purethermaliq}(c), again yields a subtler but still significant failure. While both probes lift the node under symmetric thermal bias, quantitative differences remain across a broader energy range than in the electronic case. Peak relative heat current errors approach 40\%, while remaining capped at 33.3\% at the node itself, consistent with the underlying tunnel-coupling asymmetry (see Appendix).

While the origin of the error remains unchanged, namely the finite VP heat current \( I_{\rm P,VP}^{(1)} \) and ratios of effective Onsager coefficients, its quantitative impact is reduced compared to the electronic circuit case. This is because the baseline heat current \( I_{\rm R}^{(1)} \) is generally larger in open-circuit configurations, where thermal gradients drive energy flow unobstructed by electrical bias. Consequently, the same contribution becomes less significant in relative terms, even as it remains thermodynamically inconsistent.

Finally, we examine the influence of probe type on thermopower. As shown in Fig.~\ref{fig:purethermal_thermopower}, deviations in the Seebeck coefficient \( S \) qualitatively follow the trends observed in the heat current. Under symmetric conditions, VP and VTP predictions agree closely, with errors below \( 3.6\,\mu\mathrm{V/K} \). Asymmetry introduces larger discrepancies, with peak deviations of \( 8.1\,\mu\mathrm{V/K} \) and \( 13.6\,\mu\mathrm{V/K} \) for thermal and tunnel-coupling asymmetry, corresponding to relative errors of 7.3\%, 16.3\%, and 37.1\%, respectively.


Despite these increases, thermopower remains comparatively robust. Defined under open-circuit conditions, \( S \) permits adjustment of the electrochemical potential difference \( \mu_{\rm L} - \mu_{\rm R} \), which can partially absorb inconsistencies introduced by a fixed-temperature VP. By contrast, the heat current directly reflects entropy flow and is therefore more sensitive to unphysical energy exchange, making it a more stringent diagnostic of thermodynamic consistency.

%

%
%

\section{Conclusions}

We have investigated the thermodynamic consequences of two often-used models for incorporating decoherence into quantum transport: the voltage probe (VP), which enforces local charge conservation, and the voltage–temperature probe (VTP), which additionally enforces heat conservation. While these models are often treated as equivalent, we find this to be accidental, arising only under highly symmetric conditions where the heat flows into the probe perfectly balance, a situation which is difficult to guarantee in general. Outside of such fine-tuned cases, their predictions diverge, often dramatically.

Using both analytical expressions and calculations of a realistic 1,3-benzenedithiol (1,3-BDT) molecular junction, we find that the VP model breaks down under generic non-equilibrium conditions. When thermal bias or tunnel-coupling asymmetry is introduced, the VP fails to enforce local energy balance, resulting in spurious entropy production and a lack of suppression of decoherence effects.
Most strikingly, under asymmetric thermal driving, we find the VP can fail to model decoherence at all, even with large probe coupling strengths. In the 1,3-BDT junction, we found this resulted in relative errors of four orders of magnitude in the predicted heat current. More importantly, however, this failure reflects a fundamental thermodynamic inconsistency: the VP’s fixed temperature prevents it from equilibrating with the system, violating local energy balance and rendering it incapable of mimicking physical decoherence. In contrast, the VTP enforces both charge and heat equilibration, consistently decohering interference effects as expected from a self-consistent model of dephasing.




Interestingly, some observables, such as the Seebeck coefficient, remain relatively robust to probe type. Defined under open-circuit conditions, \( S \) absorbs some inconsistencies by adjusting the electrochemical potential, rendering it less sensitive to spurious energy exchange. Nonetheless, even here, the VP introduces measurable deviations under asymmetry.



These findings underscore a fundamental point: the choice of dephasing model carries physical consequences. It encodes assumptions about local equilibration, entropy flow, and how the environment interacts with the system. While the VTP enforces thermodynamic consistency by conserving both charge and heat, the VP does not. As a result, the VP can yield qualitatively and quantitatively incorrect predictions, particularly in asymmetric or thermally driven systems. 
%
While useful in some symmetric settings, the VP should not be assumed to reliably capture decoherence effects in general nanoscale systems.

\begin{acknowledgments}
This research was graciously supported by the National Science
Foundation under award number QIS-2412920.
\end{acknowledgments}

\section*{Data Availability Statement}
The data that support the findings of
this study are available from the
corresponding author upon reasonable
request.

\appendix
\section{Persistence of the Heat Current Node in the VP Model with Asymmetric Thermal Bias}

Here we explain why the heat current \( I^{(1)}_{\rm R,VP} \) in the voltage probe (VP) model remains nearly independent of the probe coupling strength \( \Gamma_{\rm P} \) near a transmission node, even for large \( \Gamma_{\rm P} \). We consider a three-terminal system operating in linear-response under symmetric electrical bias, where
\begin{align}
T_{\rm L} &= T_0 + \delta T_{\rm L}, \quad T_{\rm R} = T_0 - \delta T_{\rm R}, \quad T_{\rm P}^{\rm VP} = T_0, \\
\mu_{\rm L} &= \mu_0 + \delta \mu, \quad \mu_{\rm R} = \mu_0 - \delta \mu.
\end{align}
The heat current into terminal \( R \) is given by
\begin{align}
I^{(1)}_{\rm R,VP} &= \Ltilde{1,1}_{\rm RL} (\mu_{\rm L} - \mu_{\rm R}) \nonumber \\
&+ \frac{1}{T_0} \left[
\Ltilde{2,1}_{\rm RL} (\delta T_{\rm L} + \delta T_{\rm R}) - \Ltilde{2,1}_{\rm RP} \delta T_{\rm R}
\right].
\end{align}

To evaluate these coefficients, we assume the transmission functions vary sufficiently slowly that it is valid to apply the Sommerfeld expansion. In the wide-band limit
\begin{align}
\mathcal{L}^{(0)}_{\alpha\beta} &\approx \frac{1}{h} \mathcal{T}_{\alpha\beta}(\mu_0), \\
\mathcal{L}^{(1)}_{\alpha\beta} &\approx \frac{\pi^2 k_B^2 T_0^2}{3h} \mathcal{T}'_{\alpha\beta}(\mu_0), \\
\mathcal{L}^{(2)}_{\alpha\beta} &\approx \frac{\pi^2 k_B^2 T_0^2}{3h} \mathcal{T}_{\alpha\beta}(\mu_0).
\end{align}
The effective Onsager coefficients can then be expressed approximately as
\begin{align}
\Ltilde{1,1}_{\alpha\beta} &\approx \frac{\pi^2 k_B^2 T_0^2}{3h} \left[
\mathcal{T}'_{\alpha\beta} - \frac{ \mathcal{T}'_{\alpha P} \, \mathcal{T}_{P\beta} }{ \mathcal{T}_{\rm PP} }
\right], \\
\Ltilde{2,1}_{\alpha\beta} &\approx \frac{\pi^2 k_B^2 T_0^2}{3h} \left[
\mathcal{T}_{\alpha\beta} - \frac{ \pi^2 k_B^2 T_0^2 }{3} \cdot \frac{ \mathcal{T}'_{\alpha P} \mathcal{T}'_{P\beta} }{ \mathcal{T}_{\rm PP} }
\right].
\end{align}


In the 1,3-benzenedithiol (1,3-BDT) junction considered in the main text, the probe is para-connected to the left electrode and ortho-connected to the right. In this configuration, the associated transmission functions \( \mathcal{T}_{\alpha P}(E) \) are nearly energy-independent in the mid-gap region.\cite{solomon2008understanding} By contrast, the direct transmission between left and right exhibits a quadratic node at mid-gap,\cite{bennett2024quantum} with \( \mathcal{T}_{\rm LR}(E) \propto E^2 \). Consequently, near the nodal energy \( \mu_0 \), both the transmission and its derivative vanish: \( \mathcal{T}_{\rm RL}(\mu_0) \approx 0 \) and \( \mathcal{T}'_{\alpha P}(\mu_0) \approx 0 \). Applying the Sommerfeld expansion, this gives
\begin{align}
	\Ltilde{1,1}_{\rm RL} &\approx 0, \\
	\Ltilde{2,1}_{\rm RL} &\approx 0, \\
	\Ltilde{2,1}_{\rm RP} &\approx \frac{\pi^2 k_B^2 T_0^2}{3h} \mathcal{T}_{\rm RP}(\mu_0).
\end{align}
Thus, the heat current reduces to
\begin{equation}
I^{(1)}_{\rm R,VP} \approx - \frac{\delta T_{\rm R}}{T_0} \cdot \frac{\pi^2 k_B^2 T_0^2}{3h} \mathcal{T}_{\rm RP}(\mu_0).
\end{equation}
This contribution is small near the node and nearly independent of \( \Gamma_{\rm P} \), as verified numerically in the main text. Note that when $\delta T_{\rm R}=0$, i.e. when $T_{\rm R}=T_{\rm P}^{\rm VP}$, the heat current is {\em independent} of $\Gamma_{\rm P}$ entirely, to first-order.


\section{Origin of the $\Gamma_{\rm P}$-Independent Relative Heat Current Correction with Asymmetric Tunnel-Coupling}

In the VP model, the probe heat current under symmetric thermal and electrical bias
\[
T_{\rm L,R} = T_0 \pm \delta T, \quad \mu_{\rm L,R} = \mu_0 \pm \delta \mu,
\]
is given by
\begin{align}
I^{(1)}_{\rm P,VP} &= \Ltilde{1,1}_{\rm PL} \delta \mu + \Ltilde{1,1}_{\rm PR} (-\delta \mu) \nonumber \\
&\quad + \frac{1}{T_0} \left[ \Ltilde{2,1}_{\rm PL} \delta T + \Ltilde{2,1}_{\rm PR} (-\delta T) \right] \\
&= \delta \mu \left( \Ltilde{1,1}_{\rm PL} - \Ltilde{1,1}_{\rm PR} \right) + \frac{\delta T}{T_0} \left( \Ltilde{2,1}_{\rm PL} - \Ltilde{2,1}_{\rm PR} \right).
\end{align}

We now evaluate these coefficients using the Sommerfeld expansion. For transmission functions that are smooth near \( \mu_0 \), we have:
\begin{align}
\mathcal{L}^{(0)}_{\alpha\beta} &\approx \frac{1}{h} \mathcal{T}_{\alpha\beta}(\mu_0), \\
\mathcal{L}^{(1)}_{\alpha\beta} &\approx \frac{\pi^2 k_B^2 T_0^2}{3h} \mathcal{T}'_{\alpha\beta}(\mu_0), \\
\mathcal{L}^{(2)}_{\alpha\beta} &\approx \frac{\pi^2 k_B^2 T_0^2}{3h} \mathcal{T}_{\alpha\beta}(\mu_0).
\end{align}

Substituting these into the effective Onsager form, and assuming \( \mathcal{T}'_{P\alpha} \approx 0 \) near the node, the cross terms in the Schur complement vanish, and we obtain
\begin{align}
\Ltilde{1,1}_{\rm P\alpha} &\approx \mathcal{L}^{(1)}_{\rm P\alpha}, \\
\Ltilde{2,1}_{\rm P\alpha} &\approx \mathcal{L}^{(2)}_{\rm P\alpha}.
\end{align}

Therefore, the probe heat current becomes
\begin{align}
I^{(1)}_{\rm P,VP} &\approx \delta \mu \left( \mathcal{L}^{(1)}_{\rm PL} - \mathcal{L}^{(1)}_{\rm PR} \right) + \frac{\delta T}{T_0} \left( \mathcal{L}^{(2)}_{\rm PL} - \mathcal{L}^{(2)}_{\rm PR} \right).
\end{align}

Near the node, \( \mathcal{L}^{(1)}_{\rm P\alpha} \sim \mathcal{T}'_{\rm P\alpha} \approx 0 \), so the electrical terms vanish and we are left with
\begin{equation}
I^{(1)}_{\rm P,VP} \approx \frac{\delta T}{T_0} \left( \mathcal{L}^{(2)}_{\rm PL} - \mathcal{L}^{(2)}_{\rm PR} \right).
\end{equation}

In the wide-band limit, the transmission from the probe to terminal \( \alpha \) behaves as
\[
\mathcal{T}_{\rm P\alpha} \propto \Gamma_{\rm P} \Gamma_\alpha,
\]
so
\[
\mathcal{L}^{(2)}_{\rm P\alpha} \propto \Gamma_{\rm P} \Gamma_\alpha.
\]

Hence, the probe heat current becomes
\begin{equation}
I^{(1)}_{\rm P,VP} \propto \Gamma_{\rm P} \left( \Gamma_{\rm L} - \Gamma_{\rm R} \right),
\end{equation}
while the heat current into the right terminal is
\begin{equation}
I^{(1)}_{\rm R,VP} = \frac{\delta T}{T_0} \Ltilde{2,1}_{\rm RP} \approx \frac{\delta T}{T_0} \mathcal{L}^{(2)}_{\rm RP} \propto \Gamma_{\rm P} \Gamma_{\rm R}.
\end{equation}

Thus, the ratio becomes
\begin{equation}
\frac{I^{(1)}_{\rm P,VP}}{I^{(1)}_{\rm R,VP}} \approx \frac{\Gamma_{\rm L} - \Gamma_{\rm R}}{\Gamma_{\rm R}}.
\end{equation}

The difference between VP and VTP models is given by~\cite{bergfield2014thermoelectric}
\begin{equation}
\delta I^{(1)}_{\rm R} = \frac{ \Ltilde{2,1}_{\rm PR} }{ \Ltilde{2,1}_{\rm PP} } I^{(1)}_{\rm P,VP}.
\end{equation}

Near the node, we may approximate
\[
\Ltilde{2,1}_{\rm PP} = -\left( \Ltilde{2,1}_{\rm PL} + \Ltilde{2,1}_{\rm PR} \right)
\approx -\left( \mathcal{L}^{(2)}_{\rm PL} + \mathcal{L}^{(2)}_{\rm PR} \right),
\]
and similarly \( \Ltilde{2,1}_{\rm PR} \approx \mathcal{L}^{(2)}_{\rm PR} \), so
\begin{equation}
\frac{\Ltilde{2,1}_{\rm PR}}{\Ltilde{2,1}_{\rm PP}} \approx -\frac{\Gamma_{\rm R}}{\Gamma_{\rm L} + \Gamma_{\rm R}}.
\end{equation}

Combining, we find the relative correction
\begin{align}
\frac{\delta I^{(1)}_{\rm R}}{I^{(1)}_{\rm R,VP}} 
&= \frac{1}{I^{(1)}_{\rm R,VP}} \cdot \frac{ \Ltilde{2,1}_{\rm PR} }{ \Ltilde{2,1}_{\rm PP} } \cdot I^{(1)}_{\rm P,VP} \\
&\approx \left( -\frac{\Gamma_{\rm R}}{\Gamma_{\rm L} + \Gamma_{\rm R}} \right) \cdot \left( \frac{ \Gamma_{\rm L} - \Gamma_{\rm R} }{ \Gamma_{\rm R} } \right) \\
&= -\frac{\Gamma_{\rm L} - \Gamma_{\rm R}}{ \Gamma_{\rm L} + \Gamma_{\rm R} }.
\end{align}

Thus, the observed peak in the relative heat current correction is a robust consequence of geometric coupling asymmetry near the node, and is independent of \( \Gamma_{\rm P} \), \( \delta \mu \), and even the precise temperature difference (provided linear response holds).

\section{Charge and Heat Currents for ``Flat'' 1,3-BDT Junction}
\label{app:flat_benzene}

\begin{figure*}
	\centering
	\subfloat[Asymmetric thermal bias; $T_{\rm L}$ = 350~K, $T_{\rm R}$ = 300~K, $T_0$ = 300~K]{
		\begin{overpic}[width=0.9\linewidth]{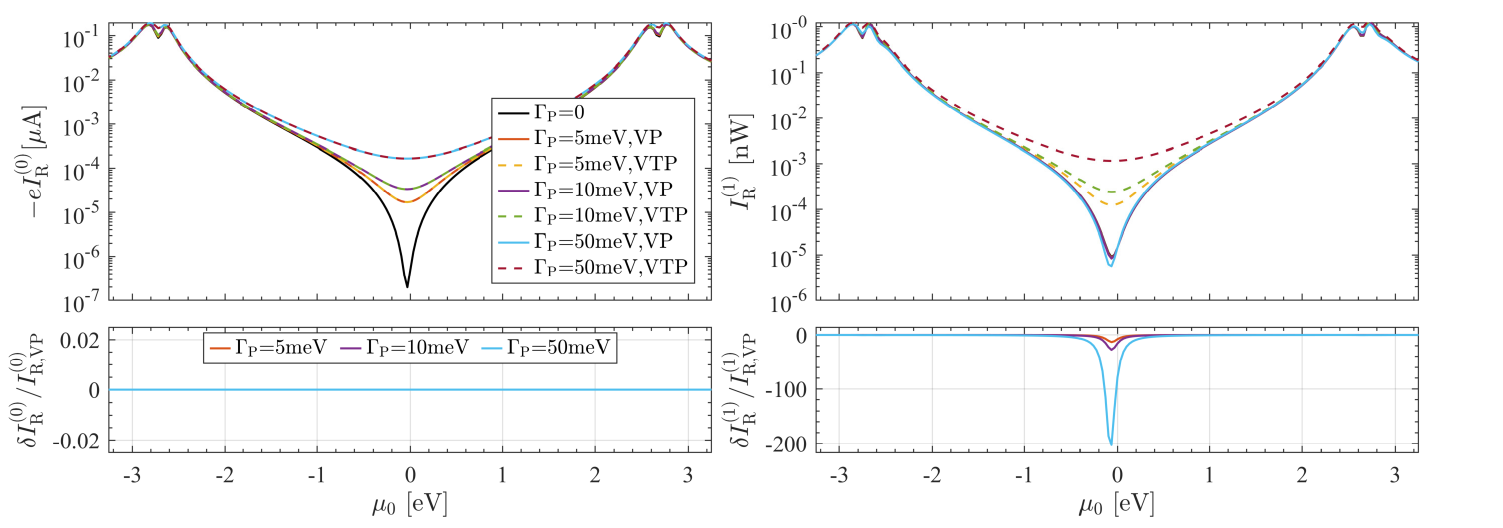}
			\put(8,16){\includegraphics[width=0.9in]{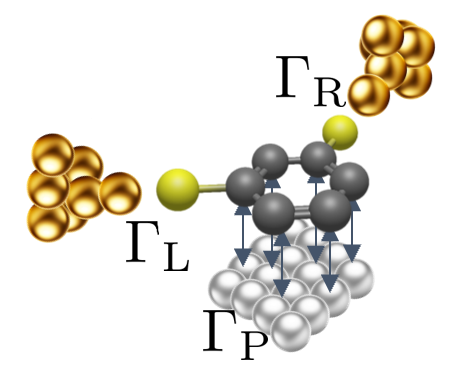}}
		\end{overpic}
	} \\
	\subfloat[Asymmetric tunnel coupling; $\Gamma_{\rm L}$ = 0.5~eV, $\Gamma_{\rm R}$ = 0.25~eV]{
		\includegraphics[width=0.9\linewidth]{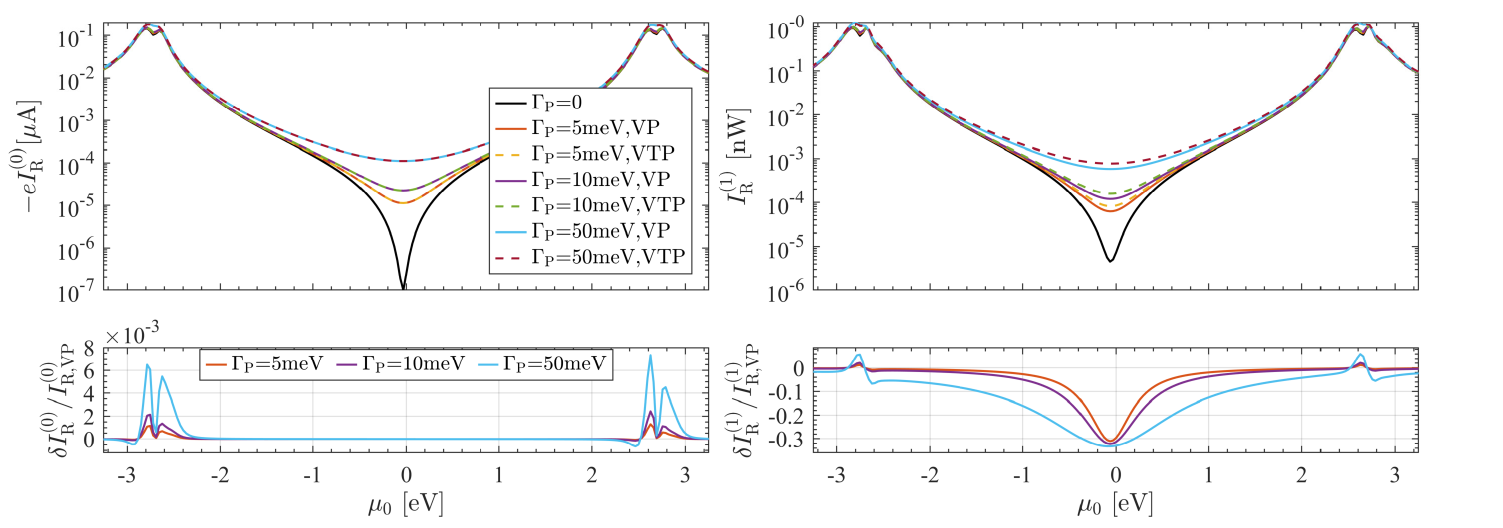}
	}
	\caption{
			Calculated charge and heat currents through a ``flat'' 1,3-BDT junction are shown for two representative sources of asymmetry and a range of probe couplings \( \Gamma_{\rm P} \), using both VP and VTP dephasing models. In all cases, a fixed electrical bias is applied ($\mu_{\rm L}$ = +5meV, $\mu_{\rm R}$=-5meV). {\bf (a)} Under asymmetric thermal bias, the charge current error vanishes identically due to complete destructive interference at the probe, with all transport amplitudes coherently mixing. The heat current error, however, is substantial, reaching $\sim$20,200\%.
		{\bf (b)} Under asymmetric tunnel coupling, the charge current error remains small, peaking near resonance at approximately 0.6\%. The heat current error again shows a robust peak at $-33.3\%$, consistent with the results of the local probe model.
	}
	\label{fig:flatbenzene}
\end{figure*}

Figure~\ref{fig:flatbenzene} shows the charge and heat currents in a 1,3-benzenedithiol (1,3-BDT) molecular junction with a flat probe configuration, in which the B\"uttiker probe couples equally to all six $\pi$ orbitals:
\begin{equation}
	\left[ \bm{\Gamma}_{\rm P} \right]_{nm} = \frac{\Gamma_{\rm P}}{6} \delta_{nm}.
\end{equation}
This coupling preserves the full cyclic symmetry of the benzene ring and ensures that the probe couples identically to all transport paths, including para-, meta-, and ortho-configured contributions.

Figure~\ref{fig:flatbenzene}(a) presents results under asymmetric thermal bias ($T_{\rm L} = 325$~K, $T_{\rm R} = 275$~K) with symmetric coupling (\( \Gamma_{\rm L} = \Gamma_{\rm R} = 0.5~\mathrm{eV} \)). Under these conditions, the charge current error vanishes identically across all probe strengths. This exact cancellation arises from the symmetric coupling of all transport amplitudes into the probe and reflects the coherent interference of para-, ortho-, and meta-configured paths. However, the heat current error reaches a dramatic peak of approximately 20,200\%, illustrating the failure of the VP model to properly account for entropy exchange under thermal asymmetry.

Figure~\ref{fig:flatbenzene}(b) shows results for asymmetric tunnel coupling (\( \Gamma_{\rm L} = 0.5~\mathrm{eV}, \Gamma_{\rm R} = 0.25~\mathrm{eV} \)) under symmetric thermal bias and fixed electrical bias ($\mu_{\rm L} = +5$~meV, $\mu_{\rm R} = -5$~meV). Here, the charge current error remains small, peaking at $\sim$0.6\% near resonance. Nonetheless, the heat current error again reaches a robust $-33.3\%$, in agreement with the behavior observed for local probes and analytic results near the node.

Finally, we note that in the ``flat'' benzene configuration, the distinction between VP and VTP models vanishes under symmetric thermal bias, tunnel coupling, and electrical bias. To understand the origin of this behavior, consider the probe heat current in the VP model under symmetric biasing conditions:
\begin{align}
	\mu_{\rm L} &= \mu_0 + \delta \mu, & \mu_{\rm R} &= \mu_0 - \delta \mu, \\
	T_{\rm L} &= T_0 + \delta T, & T_{\rm R} &= T_0 - \delta T.
\end{align}
In this case, the heat current into the VP is given by
\begin{equation}
	I_{\rm P,VP}^{(1)} = \left[ \widetilde{\mathcal{L}}^{(1,1)}_{\rm PL} - \widetilde{\mathcal{L}}^{(1,1)}_{\rm PR} \right] \delta \mu
	+ \left[ \widetilde{\mathcal{L}}^{(2,1)}_{\rm PL} - \widetilde{\mathcal{L}}^{(2,1)}_{\rm PR} \right] \frac{\delta T}{T_0}.
\end{equation}
In the flat-benzene geometry, the probe couples equally to all molecular orbitals. This uniform coupling ensures that each transmission path contributing to \( \widetilde{\mathcal{L}}^{(\nu,\xi)}_{\rm PL} \) has a mirror-symmetric counterpart contributing to \( \widetilde{\mathcal{L}}^{(\nu,\xi)}_{\rm PR} \), equal in magnitude but opposite in sign:
\begin{equation}
	\widetilde{\mathcal{L}}^{(\nu,\xi)}_{\rm PL} = -\widetilde{\mathcal{L}}^{(\nu,\xi)}_{\rm PR}.
\end{equation}
As a result, the probe heat current vanishes,
\begin{equation}
	I_{\rm P,VP}^{(1)} = 0,
\end{equation}
and the discrepancies in the charge and heat currents between VP and VTP models disappear completely.

As discussed, the exact cancellation of the heat current breaks down when the symmetry between the left and right environments is perturbed, either by asymmetric coupling or unequal temperatures. In such cases, the probe acquires a net heat flow, and the VP and VTP models yield distinct predictions for the thermal response. Remarkably, the charge current remains unaffected by thermal asymmetry. 


\bibliography{refs_clean}
\end{document}